\documentclass[aps]{revtex4}
\usepackage{amsmath,amssymb}
\usepackage{color,graphicx}
\newcommand{\be}{\begin{equation}}
\newcommand{\ee}{\end{equation}}
\begin{document}
\title{Ultrarelativistic (Cauchy)   spectral problem in the infinite well}
\author{Elena V. Kirichenko, Piotr Garbaczewski, Vladimir  Stephanovich and  Mariusz  \.{Z}aba}
\affiliation{Faculty of Mathematics, Physics and Informatics,
University of Opole, 45-052 Opole, Poland}
\date{\today }
\begin{abstract}
 We  analyze spectral properties of the ultrarelativistic (Cauchy) operator $|\Delta |^{1/2}$,
  provided its  action is  constrained exclusively to the interior of the  interval $[-1,1] \subset R$.
   To this end both analytic and numerical methods  are employed.
  New  high-accuracy spectral data are obtained.  A direct analytic proof  is given  that trigonometric functions $\cos(n\pi x/2)$ and $\sin(n\pi x)$, for integer
  $n$ are {\it not} the  eigenfunctions of $|\Delta |_D^{1/2}$,
  $D=(-1,1)$. This clearly demonstrates that the  traditional Fourier multiplier
  representation of $|\Delta |^{1/2}$ becomes defective, while passing  from $R$  to a bounded   spatial
  domain $D\subset R$.
  \end{abstract}
 \maketitle

\section{Introduction}

Fractional (L\'{e}vy-type) operators  are known to be  spatially
nonlocal. This  becomes an issue if  confronted with  a priori
imposed  exterior Dirichlet boundary data, which set the  familiar
(quantum)  infinite well  enclosure.  Standard fractional
Laplacians, at each instant of time,  extend their nonlocal action
to  the entire real axis  $R$ and this property
 needs  to be reconciled with the  finite support  $D=(-1,1)\subset R$ of the   infinite well with
 width equal $2$.

One of the obvious obstacles arising  here is  rooted in the fact, that  the  traditional Fourier
 multiplier representation of L\'{e}vy operators is no longer  operational in the finite interval,
  \cite{GS}-\cite{getoor}, see specifically \cite{pgar,zg}.  Compare e.g. also a discussion
  of  that issue for the  familiar  (Laplacian-generated) quantum mechanical infinite well problem, see e.g.
  \cite{karwowski,robinett} and \cite{zg}.

To elucidate the above  point, let us   recall that the Fourier integral
$ {\frac{1}{\sqrt{2\pi}}} \int_R   |k|^\mu \tilde{f}(k)
  e^{-\imath kx}dk  =   - \partial _{\mu }f(x)/\partial |x|^{\mu }  =  |\Delta |^{\mu
  /2}f(x)$, where $\tilde{f}$ stands for a Fourier transform of $f\in L^2(R)$ and
   $g(k)=|k|^{\mu }\tilde{f}(k)$ is presumed to be $L^2(R)$-integrable.
It is  $|k|^\mu$  which   plays the role of  the pertinent   Fourier
  multiplier.

  The above Fourier formula  is  quite  often  interpreted  as a  universal   definition of   both the
  fractional operator $- |\Delta |^{\mu /2}$ and that of the
     fractional derivative of the $\mu $-th order,   $\partial _{\mu }f(x)/\partial |x|^{\mu }  =-|\Delta |^{\mu  /2}f(x)$,
  for $\mu \in  (0,2)$. However, this definition is  unquestionably  valid  only,  if the fractional operator is  defined on  the whole real line
   $R$. More than that, it   appears to be merely  a specific  admissible choice  in the family of equivalent (while on $R$)  definitions, \cite{ten}.

  The nonlocal operator  $- |\Delta |^{\mu /2}$,
  is known to generate  two versions   of so-called  fractional dynamics
  (dimensional constants being scaled away):
 (i) the   semigroup $\exp(- t |\Delta |^{\mu /2}) \,f$ and
 (ii)  unitary  $\exp(- it |\Delta |^{\mu /2})\, f$ ones.

Apart from the unperturbed (free)  case,  the Fourier  (multiplier) representation of the fractional dynamics
 has proved useful if an infinite or periodic support is  admitted for functions in the  domain, \cite{ZRK}.
  For the  simplest quadratic ($\sim x^2$) perturbation of the fractional Laplacian (the  fractional oscillator problem),
  a complete analytic solution  has been found   in the ultrarelativistic (Cauchy) oscillator  case,
   by resorting to Fourier space (and Fourier multiplier)  methods.

 For more complicated  perturbations, and likewise for  a deceivingly  simple problem of the fractional Laplacian  in a bounded
  (spatial)  domain,  standard  Fourier techniques  have been found  to be  of a doubtful or   limited use, \cite{ZRK}.
  Therefore, to keep spatial constraints under control,  we turn over to  a fully-fledged  spatially nonlocal    definition of the   fractional  Laplacian, that is  well known in the
  mathematical and statistical physics literature  \cite{ZG}-\cite{K}, while seldom  invoked by quantum theory practitioners, see however \cite{GS}  and references
   there in.

Dating back to the classic papers  \cite{riesz,hadamard}, one interprets the  fractional Laplacian $-|\Delta |^{\mu /2}$,  $\mu \in (0,2)$ as
a pseudo-differential  (integral)   spatially nonlocal  operator and its  action   on a
function from  the   $L^2(R)$  domain  is  commonly  defined     by  employing   the Cauchy
   principal value  of the  involved  integral (evaluated relative to  singular points of integrands)
   \begin{equation}
 |\Delta |^{\mu /2} f(x)\, =\,
 -  {\frac{\Gamma (\mu +1) \sin(\pi \mu/2)}{\pi }} \int_R {\frac{f(z)- f(x)}{|z-x|^{1+\mu }}}\, \label{integral}
 dz .
\end{equation}
For a rationale and  a broader discussion of the  uses (and misuses) of  this formula, including its Fourier multiplier version,
 see  \cite{GS}.

By departing from the general   spatially nonlocal definition (\ref{integral}) we shall  pass to  the specialized Cauchy case ($\mu
=1$) of the fractional Laplacian  and next   focus our attention on  its
properties under   exterior Dirichlet boundary data (e.g. the infinite well enclosure).   This  issue has
received some  coverage in the literature, both physics-oriented
\cite{ZG}-\cite{ZRK} and purely mathematical
\cite{getoor}-\cite{dyda}. See also Ref. \cite{GS} for additional
references and a discussion of earlier attempts to   find   the
spectral  solution for  the  infinite  fractional  well.

In the present paper, the  term  "ultrarelativistic" directly  stems form the notion of the
 quasi-relativistic operator $\sqrt{- \Delta + m^2}$  (natural units being  presumed)
 and its mass $m\rightarrow 0$ limit $|\Delta |^{1/2}$, see e.g. \cite{GS}  and \cite{rel}.

\section{The infinite well enclosure: From $|\Delta |^{1/2}$  to $|\Delta |_D^{1/2}$.}

The  Hamiltonian-type expression $H=-  |\Delta |^{1/2} + V$, with $V(x)=0$ for $x\in D=(-1,1)\subset R$,
   is an  encoding    of the Cauchy operator with the   Dirichlet  boundary conditions  (so-called  zero exterior
   condition on $R\setminus D$)  imposed on  $L^2(R)$  functions $f(x)$  in  the domain of $H$:   $f(x) =0$ for $|x| \geq 1$.
We point out that the Cauchy operator  $|\Delta |^{1/2}$ if restricted to a domain comprising solely  $L^2(R)$ functions
with a support in $D$ and vanishing on $R\backslash D$   is not a self-adjoint  operator in $L^2(R)$.

 However, if we  consider  the action of  $|\Delta |^{1/2}$   on test functions    $f\in C_0^{\infty }(D)$ (infinitely differentiable functions
  that are compactly supported in $R$ ),   then  the restriction $|\Delta |^{1/2}_D$ of
$|\Delta |^{1/2}$  to  $D$ is interpreted as the Cauchy operator with the  zero (Dirichlet)  exterior condition
on $R\backslash D$   and is known to extend to a self-adjoint operator in $L^2(D)$, \cite{K}.
The passage from $C_0^{\infty }(R)$ to $C_0^{\infty }(D)$  ultimately  amounts to disregarding   any  $R\backslash D$
 contribution implicit in the formal definition   (\ref{integral}).

  Let us discuss  the  $D$ versus $R\backslash D$ interplay in more detail, by  considering  the action of $|\Delta |^{1/2}$
  on   these  $C_0^{\infty }(R)$  functions which are actually supported in $D$, i.e. $\psi \in  C_0^{\infty }(D)$,
   while departing from the original nonlocal definition:
\begin{equation} \label{stu1}
|\Delta |^{1/2}\psi(x)=-\frac{1}{\pi}\int_R\frac{\psi(x+y)-\psi(x)}{y^2}dy.
\end{equation}

\begin{equation} \label{st1}
\psi(x)=\left\{\begin{array}{c}\psi(x), \ x\in (-1,1) \\0, \quad {\rm {otherwise}},
\end{array}\right.
\end{equation}
Given $x\in (-1,1)$,  we realize that
$\psi(x+y)$ does not  vanish   identically if $ x+y\in (-1,1)$   i.e.  for $ -1-x < y < 1-x$.
Therefore,  the integration (\ref{stu1}) can be
simplified  by  decomposing  $R$ into $(-\infty <y \leq  -1-x) \cup (-1-x<y<1-x) \cup (1-x \leq y <\infty )$.
We have:
\begin{eqnarray}
|\Delta |^{1/2}\psi=-\frac{1}{\pi} \left[-\psi(x)\left(\int_{-\infty}^{-1-x}\frac{dy}{y^2}+\int_{1-x}^\infty\frac{dy}{y^2}\right)+\int_{-1-x}^{1-x}\frac{\psi(x+y)-\psi(x)}{y^2}dy\right]=\nonumber \\
=\frac{2}{\pi}\frac{\psi(x)}{1-x^2}-\frac{1}{\pi}\int_{-1-x}^{1-x}\frac{\psi(x+y)-\psi(x)}{y^2}dy, \label{stu2}
\end{eqnarray}
where the second integral  should be understood as the  Cauchy principal value with respect to $0$, i.e. $
\int_{-1-x}^{1-x}=\lim_{\varepsilon \to 0} \left[\int_{-1-x}^{-\varepsilon} +\int_{-\varepsilon}^{1-x}\right]$.

Given $x\in (-1,1)$, let us make a substitution $x+y=t$ in \eqref{stu2}, presuming that now  the Cauchy principal value needs to
 be evaluated relative to $x$. We obtain (note the   principal value  $(p.v.)$ symbol, introduced  in the self-explanatory notation)
\begin{eqnarray}
&&|\Delta |^{1/2}\psi=\frac{2}{\pi}\frac{\psi(x)}{1-x^2}+\frac{1}{\pi}   \int_{-1}^{1}\frac{\psi(x)-\psi(t)}{(t-x)^2}dt=
\frac{2}{\pi}\frac{\psi(x)}{1-x^2}+  \frac{1}{\pi}(p.v.) \left[\left.-\frac{\psi(x)}{t-x}\right|_{-1}^1
-\int_{-1}^{1}\frac{\psi(t)dt}{(t-x)^2}\right]=\nonumber \\
&&=\frac{1}{\pi}\, \lim_{\epsilon \rightarrow 0} \left[ {\frac{2\psi (x)}{\epsilon }}  -
\int_{-1}^{{x-\epsilon }}\frac{\psi(t)dt}{(t-x)^2}
 -\int_{x+\epsilon }^{1}{\frac{\psi(t)dt}{(t-x)^2}} \right]
   \equiv -\frac{1}{\pi} ({\cal{H}}) \int_{-1}^{1}\frac{\psi(t)dt}{(t-x)^2}, \label{ii1}
\end{eqnarray}
where $({\cal{H}})$ refers to the  Hadamard regularization of hypersingular integrals (Hadamard finite part, extensively employed in
the engineering literature, \cite{hadamard}-\cite{klerk}). We  point out that the troublesome term
$\frac{2}{\pi}\frac{\psi(x)}{1-x^2}$ has been cancelled  away by its negative coming from the evaluation
of $(p.v.)[...]$ in the above.

The second line of the formula (\ref{ii1})  can be interpreted as a
 definition   of    $|\Delta |^{1/2}_D$.  The pertinent operator,  instead of referring   merely  to $C_0^{\infty }(D)$ functions,
can be literally applied (extended)  to functions  $\psi \in L^2(D)$.

In the literature on the usage of the Hadamard finite part evaluation of hypersingular integrals, it is often mentioned
that if the $({\cal{H}})$
 integral and the $(p.v.)$ integrals in question do exist, we can relate  them as follows:
\begin{equation}
 |\Delta |^{1/2}_D \psi (x) = -\frac{1}{\pi}\, ({\cal{H}}) \int_{-1}^{1}\frac{\psi(t)dt}{(t-x)^2}= -
 \frac{1}{\pi}\, {\frac{d}{dx}} (p.v.) \int_{-1}^{1}{\frac{\psi(t)dt}{t-x}}. \label{ii2}
\end{equation}
We shall  employ another version of the  Hadamard - Cauchy integral relation,  by  following a direct  integration by parts
 procedure and continually  keeping in mind that the involved integrals are (hyper)singular, c.f. also
 \cite{klerk,monegato}.

Namely, by invoking the second line of Eq. (\ref{ii1}) and performing integrations by parts
 before the limit $\epsilon \rightarrow 0$ is ultimately taken, we end up  with:
\be \label{7}
|\Delta |^{1/2}_D \psi (x) =  - \frac{1}{\pi}\, ({\cal{H}}) \int_{-1}^{1}\frac{\psi(t)dt}{(t-x)^2}=
 - \frac{1}{\pi}\, (p.v.)\int_{-1}^{1}\frac{\psi '(t)dt}{t-x},
\ee
where $\psi '(t)= d\psi (t)/dt$.

We are  interested in solving an  eigenvalue problem  $|\Delta |^{1/2}_D \psi = E\, \psi $  for the infinite Cauchy well, while
 interpreted in terms of the   hypersingular integral equation.  For explicit computations
 we shall   employ  the Hadamard (finite part)-Cauchy (principal value)  relation  (\ref{7}):
\be  \label{8}
 E\, \psi (x)  +  \frac{1}{\pi}\, (p.v.)\int_{-1}^{1}\frac{\psi '(t)dt}{t-x} = 0 .
\ee

\section{$\cos(\pi x/2)$ and   $\sin(\pi x)$  are not  eigenfunctions  of $|\Delta |^{1/2}_D$.}

In Ref. \cite{GS} we have discussed  the validity of counter-arguments  against proposed so-far, in the physical literature,
spectral solutions for L\'{e}vy-stable  infinite well problems  \cite{jeng,luchko}.
  By invoking rigorous mathematical results  of \cite{K,KKMS} we  have given in \cite{ZG,zg}
 the  computer-assisted proofs (elaborated for the infinite Cauchy well)  that
 spectral results of  \cite{laskin}-\cite{iomin} are surely  incorrect
    in  the lower  part of the spectrum and may be employed at most as  approximate expressions for higher  eigenvalues.
    In particular a computer-assisted  analysis of approximate eigenfunctions shapes   \cite{zg} have
    demonstrated quite clearly that "plain" trigonometric functions
    (like e.g. sine  and cosine) are not the eigenfunctions for the problem under consideration, see also \cite{luchko}.

   Presently we shall demonstrate  analytically that  $\cos (\pi x/2)$ is {\it not} the ground state function of
    $|\Delta |^{1/2}_D$,  so  contradicting  claims   of  \cite{bayin,bayin1,iomin}.  Our method is different from that adopted  in Ref \cite{luchko}.
 The  integrations are  to be performed in the Hadamard sense, c.f. (\ref{ii1}) and that  will allow us to
 introduce basic tools  that will be necessary in the  subsequent  more general   spectral analysis.

Let us directly substitute $\psi (x) = \cos(\pi x/2)$  to  Eq.(\ref{ii2}). We shall demonstrate that:
\begin{equation} \label{las5}
 |\Delta |^{1/2}_D \cos\frac{\pi x}{2}=-\frac{1}{\pi}  ({\cal{H}})  \int_{-1}^{1}\frac{\cos\frac{\pi t}{2}dt}{(t-x)^2}=
\frac 12 \cos\frac{\pi x}{2}\left[{\mathrm {Si}}\frac{\pi(1+x)}{2}+{\mathrm {Si}}\frac{\pi(1-x)}{2}\right]+
\frac 12 \sin\frac{\pi x}{2}\left[{\mathrm {Ci}}\frac{\pi(1-x)}{2}-{\mathrm {Ci}}\frac{\pi(1+x)}{2}\right],
\end{equation}
which surely remains incompatible with any function of the form   $ E\, \cos(\pi x/2)$, where  $E>0$ is a constant and $x\in (-1,1)$.
Here  ${\mathrm{Ci}}(x)$    and ${\mathrm{Si}}(x)$  are respectively   the cosine  and sine integral  functions,
which  are defined as follows, \cite{abr}:
\begin{equation} \label{las1a}
{\mathrm   {Si}}(x)=\int_0^x\frac{\sin t}{t}dt= \frac{1}{\pi }  - \int_x^{\infty } \frac{\sin t}{t} dt
\end{equation}
is an entire function on $R$ with the properties $ {\mathrm {Si}}(\infty )= \pi /2 $ and ${\mathrm {Si}}(-x) = - {\mathrm  {Si}}(x)$,
  while
\be \label{las1b}
{\mathrm   {Ci}}(x)= - \int_x^{\infty } \frac{\cos t}{t}dt  = C + \ln x  + \int_0^x \frac{\cos t -1}{t}dt
\ee
is restricted to $R^+$and  $C= - \int _0^{\infty } e^{-t} \ln t \, dt = 0.577215665...$ stands for the Euler-Mascheroni constant.

 We point out that  ${\mathrm  {Si}}(x)$ is defined everywhere on $R$ as a continuous and differentiable function. On the contrary,
${\mathrm   {Ci}}(x)$,  $x\in R^+$   logarithmically  escapes
 towards $-\infty $  as $x$ drops down to $0$ \cite{abr,GR}.
    In the vicinity of the well boundaries  $x \to \pm 1$ of $D$,
 the   logarithmic divergence  $\ln(1-|x|) \to - \infty $    definitely dominates.

For the direct evaluation of   $|\Delta |^{1/2}_D \cos\frac{\pi x}{2}$, with $x\in (-1,1)$ we shall employ the Cauchy  principal value formula (\ref{7}).
According to Eq. (\ref{7}), we have:
\be
|\Delta |^{1/2}_D \cos(\frac{\pi x}{2})= {\frac{1}{2}}  (p.v.) \int_{-1}^{1} \frac{\sin(\pi t/2)}{t-x}  dt =
{\frac{1}{2}}  (p.v.) \int_{-1-x}^{1-x} \frac{\sin[\pi (u+x)/2]}{u}  du=
\ee
$$
+{\frac{1}{2}} \cos(\pi x/2)\, (p.v.) \int_{-1-x}^{1-x} {\frac{\sin(\pi u/2)}{u}}  du      +
 {\frac{1}{2}}    \sin(\pi x/2)  (p.v.) \int_{-1-x}^{1-x} {\frac{\cos(\pi u/2)}{u}}  du.
$$
By employing the definition (\ref{las1a}), we readily get
 \be (p.v.)
\int_{-1-x}^{1-x} {\frac{\sin(\pi u/2)}{u}}  du = \lim_{\epsilon
\downarrow 0} \left[\int_{-1-x}^{-\epsilon }   + \int^{1-x}_{\epsilon
}\right]{\frac{\sin(\pi u/2)}{u}}  du = {\mathrm   {Si}}{\frac{\pi
(1+x)}{2}}  +{\mathrm   {Si}}{\frac{\pi (1-x)}{2}}. \ee
 To evaluate  $(p.v.) \int_{-1-x}^{1-x} {\frac{\cos(\pi u/2)}{u}}  du$, let us notice that
 \be
(p.v.) \int_{-1-x}^{1-x} {\frac{\cos(\pi u/2)   -1}{u}}  du= \lim_{\epsilon
\downarrow 0} \left[\int_{-1-x}^{-\epsilon }   + \int^{1-x}_{\epsilon
}\right] {\frac{\cos(\pi u/2)   -1}{u}}  du =\ln {\frac{1+x}{1-x}} +  \left[\int_0^{1+x} -
 \int_0^{1-x}\right]{\frac{\cos(\pi u/2)}{u}}  du.
 \ee
By employing (\ref{las1b}), we  get
\be
(p.v.) \int_{-1-x}^{1-x} {\frac{\cos(\pi u/2)}{u}}  du = {\mathrm   {Ci}}(1-x) - {\mathrm   {Ci}}(1+x)
\ee
and the identity (\ref{las5})  readily  follows.  Clearly, the outcome  (\ref{las5}) is incompatible with  $E\, \cos (\pi x/2)$,
$E>0$, as  predicted in \cite{laskin}, \cite{bayin,bayin1,iomin})

With all  necessary tools in hands, one can easily verify that the the lowest odd would-be (candidate) eigenfunction $\sin(\pi x)$
(that according to \cite{laskin}, \cite{bayin,bayin1,iomin}) of $|\Delta |^{1/2}_D$ is a faulty guess.
Namely, we have:
\be
|\Delta |^{1/2}_D \sin(\pi x)=-\frac{1}{\pi} ({\cal{H}})\int_{-1}^1\frac{\sin \pi t}{(t-x)^2}dt=
\sin (\pi x)\Bigl({\rm{Si}}[ \pi (1-x)] +{\rm{Si}}[\pi (1+x)]\Bigr)
-\cos (\pi x)\Bigl({\rm{Ci}}[\pi (1-x)] - {\rm{Ci}}[\pi (1+x)]\Bigr),
\ee
while an expected  outcome  (according to \cite{laskin}, \cite{bayin,bayin1,iomin}) should be $E' \, \sin (\pi x)$, where  $E' >0$ is a constant.
This is definitely not the case.  We point out that a logarithmic divergence becomes dominant at the   boundaries of
 the interval $D$,  that  in view of
 ${\rm{Ci}}[\pi (1-x)]\rightarrow  -\infty $ for  $x\uparrow 1$  and $- {\rm{Ci}}[\pi (1+x)] \rightarrow +\infty $
 as $ x\downarrow -1$.

The above discussion  easily extends to   more general formulas  (\ref{so6}) and (\ref{so23}) in below,  which provide
a direct analytic demonstration that    trigonometric
functions of the form  $\cos(n\pi x/2)$ and $\sin(n\pi x/2)$ with $n$ integer, are {\it not} the eigenfunctions of
 $|\Delta |^{1/2}_D$. That  invalidates  claims to the contrary,   appearing in the literature on so-called fractional
 quantum mechanics,  \cite{laskin}, \cite{bayin,bayin1,iomin}.

 We note that on formal grounds, the trigonometric functions seem to be valid eigenfunctions if the Fourier  multiplier representation (c.f. Section I)
  is "blindly" used, ignoring the subtleties  related to  Fourier integrals of functions with  support in a bounded domain, \cite{ZG,karwowski,robinett}.
   The point is that the primary, mathematically well founded,   definition of the fractional operator is  provided by  the integral formula
    (\ref{integral}) and not by  its Fourier integral version. The latter is merely  a derived one  while on $R$,  \cite{ten}.
 If spatial constraints are imposed, we may  keep their effects under tight   control  only  on the level of Eq. (\ref{integral}), see our
 considerations of Section II.

\section{Solution of the eigenvalue problem in the infinite ultrarelativistic  well}

Now we are going to solve the integral equation \eqref{8}, i.e. to deduce the  eigenfunctions and eigenvalues  of the nonlocal operator $|\Delta |^{1/2}_D$.
 We note \cite{pol} that there are no  worked out   systematic   methods (even numerical)  of   solution of integral equations if their kernels are singular, or (that is worse)
   hypersingular.
 In below  we shall  provide an example of a successful solution  method,  based on Fourier series  (trigonometric)
 expansion in $L^2(D)$.   Derivations of approximate eigenfunctions and eigenvalues
 are  computer-assisted. The  outcomes   converge  slowly   towards  "true" solutions due to the singular behavior of
 ${\rm{Ci}}$ at the boundaries of $D$.

To find the eigenfunctions and eigenvalues   of the nonlocal  operator $|\Delta |^{1/2}_D$, we adopt  the  following assumptions:
\begin{itemize}
\item[1.] Based on standard quantum mechanical (Laplacian based) infinite well experience and  previous attempts,   \cite{K,KKMS} and
 \cite{ZG}-\cite{zg},  to solve the L\'{e}vy-stable    infinite   well  problem   we can safely  classify  eigenfunctions  to be odd or  even.
The  oscillation theorem appears here to be valid and the ground state has no nodes (intersections with $x$ axis), first excited state has one node, second one has two nodes etc.
So, our even states can be labeled by quantum numbers $k=$ 0,2,4,6,... while odd states by  $k=$1,3,5,....
\item[2.] The (Hilbert)  state space of the system can be interpreted as a direct sum  of odd and even (sub)spaces, equipped with basis systems comprising respectively  even and odd orthonormal sets of
functions in the interval [-1,1].
\item[3.]  In accordance with   the  infinite well  boundary conditions, the   function  in the domain of  $|\Delta |^{1/2}_D$
must obey  $\psi(x )=0$ for $|x|\geq 1$. In consequence, among various orthogonal sets  available in  $L^2(D)$, we are ultimately left with standard trigonometric
 functions.
\item[4.] The  even basis system in $L^2(D)$ is composed of cosines
\begin{equation}\label{so1}
\varphi_k(x)=\cos \frac{(2k+1)\pi x}{2},\quad  \int_{-1}^1\varphi_k(x)\varphi_l(x)dx=\delta_{kl},
\quad  k\geq 0,
\end{equation}
where $\delta_{kl}$ is the  Kronecker symbol.
 For the odd  basis system  we  take  the sines
\begin{equation}\label{so2}
\chi_k(x)=\sin k\pi x,\quad   \int_{-1}^1\chi_k(x)\chi_l(x)dx=\delta_{kl},\quad  k\geq 1.
\end{equation}

\item[4.] We look for eigenfunctions of  $|\Delta |^{1/2}_D$  separately in  odd and even Hilbert (sub)spaces  of $L^2(D)$.

 Presuming that the Fourier (trigonometric)  series converge, for even functions we have
\begin{equation}\label{so3}
\psi_e(x)=\sum_{k=0}^\infty a_k\cos \frac{(2k+1)\pi x}{2},
\end{equation}
while for odd functions
\begin{equation}\label{so3a}
\psi_o(x)=\sum_{k=1}^\infty b_k\sin k\pi x.
\end{equation}
\end{itemize}

To avoid confusion, we point out that the standard numbering of   overall  infinite well  eigenfunctions   begins from $n=1$ rather then from  $k=0$  (even case) or $k=1$ (odd case) as we  have assumed above.
We need to have  a clear discrimination between    sine (odd) and cosine (even) Fourier series expansions.  The final outcomes will be re-labeled according to the traditional lore, i.e. in terms of consecutive
 integers $n=1,2,...$.

\subsection{Even subspace}

In the present case  we substitute the function $\psi_e(x)$ \eqref{so3} into  \eqref{8} to obtain
\begin{equation}\label{so4}
\sum_{k=0}^\infty a_k f_k(x)=E \sum_{k=0}^\infty a_k \cos \frac{(2k+1)\pi x}{2},
\end{equation}
where
\begin{eqnarray}
f_k(x)&=&-\frac{1}{\pi}({\cal{H}}) \int_{-1}^1\frac{\cos \frac{(2k+1)\pi t}{2}}{(t-x)^2}dt=\label{so5} \\
&=&\frac{1+2k}{2}\Biggl\{\sin \frac{(2k+1)\pi x}{2}\left[{\rm{Ci}}\frac{(2k+1)\pi (1-x)}{2}-
{\rm{Ci}}\frac{(2k+1)\pi (1+x)}{2}\right] +\nonumber \\
&+&\cos \frac{(2k+1)\pi x}{2}\left[{\rm{Si}}\frac{(2k+1)\pi (1-x)}{2}+
{\rm{Si}}\frac{(2k+1)\pi (1+x)}{2}\right] \Biggr\}.\label{so6}
\end{eqnarray}

We recall  that the functions ${\rm Ci}[\frac{(2k+1)\pi (1+x)}{2}]$ are singular at $x \to \pm 1$ and diverge as  $\ln(1-|x|)$.
Nonetheless,  matrix elements  computed in below  prove to be  finite.
It is the  singularity of $f_k(x)$ which slows down a  convergence   of  approximate  expressions
 for  $|\Delta |^{1/2}_D\psi_e(x)$ (finite series expansions of  increasing accuracy)
 to the corresponding  "true" eigenfunctions $E \psi_e(x)$.

Let us  multiply both sides of the  equation \eqref{so4} by $\varphi_i(x)$ \eqref{so1}  and   integrate from $-1$ to $1$,   while employing
the  orthonormality   of   $\varphi_i(x)$.  The  equation  \eqref{so4} is now replaced by an (infinite) matrix eigenvalue problem
\begin{equation}\label{so7}
\sum_{i,k=0}^\infty a_k \gamma_{ki}=E a_l,\quad   \gamma_{ki}=\int_{-1}^1f_k(x)\varphi_i(x)dx,\quad  i,k,l=0,1,2,3,...
\end{equation}
whose solution will be sought for  in terms of   a sequence of   eigenvalue problems for   finite $n\times n$   matrices,
 with a gradually increasing  degree  $n$.

The set \eqref{so7} is the linear homogeneous system, which, according to Kronecker-Capelli theorem,  has a nontrivial solution  only if its determinant equals zero.
This permits to determine  the eigenvalues $E_k$ and the coefficients $a_k$ of the expansion \eqref{so1} as the eigenvectors, corresponding to each $E_k$.  That separately for each degree $n$ of the involved matrix.

While solving  Eq. \eqref{so7} numerically, the best way to calculate $\gamma_{ki}$ is computer-assisted as well  (integrals necessary to evaluate  $\gamma_{ki}$ are no more divergent,
 so that their numerical calculation is straightforward), but it turns out that a number of them
  can be computed  analytically. The analytical calculation permits to establish the fact that
  the matrix \eqref{so7} is symmetric, i.e. $\gamma_{ki}=$ $\gamma_{ik}$ and thus the sought for  eigenvalues  are real.

  In particular, we have
   \begin{equation}\label{so11}
\gamma_{kk}=-\frac{2}{\pi}+(2k+1){\rm{Si}}[\pi(2k+1)].
\end{equation}
Some  exemplary    $\gamma_{ki}$ are worth reproducing as well:
\begin{eqnarray}
&&\gamma_{00}=-\frac{2}{\pi}+{\rm{Si}}(\pi)=1.21531728,\quad   \gamma_{10}=\gamma_{01}=
\frac{6{\rm{Ci}}(\pi)-6{\rm{Ci}}(3\pi)+\ln 729}{8\pi}=0.2773259,\nonumber \\
&&\gamma_{20}=\gamma_{02}=-\frac{5}{24\pi}\left(2{\rm{Ci}}(\pi)-2{\rm{Ci}}(5\pi)+\ln 25\right)=-0.2227035, \nonumber \\
&&\gamma_{21}=\gamma_{12}=\frac{5}{16\pi}\left(6{\rm{Ci}}(3\pi)-6{\rm{Ci}}(5\pi)+
\ln\frac{15625}{729}\right)=0.3088509. \label{so12}
\end{eqnarray}

{\bf Remark:} For the reader's convenience, let us mention that an analytic evaluation of matrix elements can be greatly simplified by taking advantage of
worked out indefinite integral formulas (Section 5.3  of \cite{GR}) for e.g.
 $\int \cos(\alpha x) {\rm Ci}(\beta x) dx$,  $\int \sin(\alpha x) {\rm Ci}(\beta x)dx$ and analogous integrals with ${\rm Ci}$
  replaced by ${\rm Si}$. It is worthwhile to notice that if such  integrals   contain  products of trigonometric functions instead of "plain"
  ones, we can always  reduce them  to  one of the listed forms by employing   various trigonometric identities. Example:
  $2 \sin(\alpha x)\cos (\gamma x) = \sin[(\alpha + \gamma )x] + \sin[(\alpha - \gamma )x]$.\\

An  explicit form of the matrix \eqref{so7}, once  we truncate the infinite series at a finite $n$, reads
\begin{equation}\label{so13}
\hat A_D=\left(\begin{array}{cccc}\gamma_{00} & \gamma_{10} & \cdots & \gamma_{n0} \\
\gamma_{10} &\gamma_{11} & \cdots & \gamma_{n1}  \\
 \vdots & \cdots & \cdots & \vdots \\
\gamma_{n0} & \gamma_{n1}  & \cdots & \gamma_{nn}
\end{array}\right).
\end{equation}
To find its eigenvalues and eigenvectors we use iterative procedure, considering partial matrices $2\times 2$, $3\times 3$
 etc. The eigenvalues of  the  simplest partial matrix $2\times 2$ give the lowest order approximation of  ground state
 and {\em{second}} excited state $n=2$. The equation for associated
  eigenvalues reads:
\begin{equation}\label{so14}
\left|\begin{array}{cc} \gamma_{00}-E& \gamma_{10} \\
\gamma_{10} & \gamma_{11}-E
\end{array}\right|=0,
\end{equation}
The analytical expressions for $E_0$ and $E_2$ can be obtained  by means  of analytical formulas for $\gamma_{ik}$ \eqref{so12}.
Although  computations are  cumbersome,  one arrives at  a reasonable (albeit still far form being sharp)
 approximation to  eigenvalues associated with  the ground state and first  (even)
 excited state.
 Using numerical values of
$\gamma_{ik}$ \eqref{so12}, we deduce
\begin{eqnarray}
E_0=1,191256,\ E_2=4.411727 \ - \ {\rm{eigenvalues}},  \label{so15} \\
\psi (E_0)=(-0.996257,\ 0.086437), \ \psi (E_2)=(0.086437,\ 0.996257)
\ - \ {\rm{eigenvectors}}. \label{so16}.
\end{eqnarray}
In other words, the approximate (crude, low order) shapes  of the eigenfunctions read
\begin{eqnarray}
\psi_0=-0.996257\cos \frac{\pi x}{2}+0.086437\cos \frac{3\pi x}{2}
\ - \ {\rm{ground\ state}},  \label{so17} \\
\psi_2=0.086437\cos \frac{\pi x}{2}+0.996257\cos \frac{3\pi x}{2}  \label{so18}
\ - \ {\rm{second\ excited\ state}}.
\end{eqnarray}
We note here that the reproduced  eigenvectors are $L^2(D)$ normalized, while an overall sign may be negative, which is  immaterial for the validity of the spectral solution.

By increasing the matrix order from $2$ to $3$, we improve the accuracy  with which  lowest states are reproduced and
increase their number by one.  We have for eigenenergies
\begin{equation}\label{so19}
E_0=1.1814891,\ E_2=4.3854565,\ E_4=7.569241.
\end{equation}
It is seen that while one more state appears,  numerical outcomes for  lowest states
are  corrected  by approximately 1\%.

For  the  $6\times 6$ matrix we have
\begin{equation}\label{so20}
E_0=1.1704897,\ E_2=4.35648331,\ E_4=7.52132, \ E_6=10.68291, \ E_8=13.845025,\ E_{10}=17.01393.
\end{equation}
We note that   the value $E_0$ \eqref{so20}  is quite close to the  (still crude)   approximate eigenvalue
 $E_{gs}=  3\pi/8=1.1781$  deduced  in  Ref. (\cite{KKMS,K}). According to \cite{K} the
  infinite Cauchy well eigenvalues $E_n$   become close to  $(\frac{n\pi }{2} - \frac{\pi }{8})
  \rightarrow \frac{n\pi }{2}$, as $n\rightarrow \infty $.
  Obviously, while passing to higher order matrices the obtained eigen-solutions  give  better approximations of
  "true" eigenvalues  and eigenvectors in the infinite Cauchy well problem.

The analysis of numerical values of matrix   elements in  \eqref{so13} shows that these of diagonal elements are much larger than the  off-diagonal ones.
This difference appears to be lowest for  $\gamma_{00}$ which equals $1.215$, while off-diagonal elements take  values around 0.3, see \eqref{so12}.
For larger $k$ the diagonal elements grow (for example $\gamma_{22}\approx 4.388$), while off-diagonal  values   remain  close to  0.3.
This means  that  diagonal elements (expression \eqref{so11} for even states and \eqref{so24} for odd ones) give
a fairly  good (even if crude)  approximation for eigenvalues of the matrix \eqref{so13}.  Compare e.g. also the first row of Table I.

\subsection{Odd subspace}

We look for  eigenfunctions  in the form \eqref{so3a}. Repeating  the same steps as for  the  even subspace we generate
the following set of equations
\begin{eqnarray}
&&\sum_{i,k=1}^\infty b_k \eta_{ki}=E b_l,\quad \,  \eta_{ki}=\int_{-1}^1g_k(x)\chi_i(x)dx,\ i,k,l=1,2,3,..., \label{so21}  \\
g_k(x)&=&-\frac{1}{\pi}\int_{-1}^1\frac{\sin k\pi t}{(t-x)^2}dt=  \nonumber \\
&=&k\Bigl\{\sin (k\pi x)\Bigl({\rm{Si}}[k\pi (1-x)]+{\rm{Si}}[k\pi (1+x)]\Bigr)
-\cos (k\pi x)\Bigl({\rm{Ci}}[k\pi (1-x)]- {\rm{Ci}}[k\pi (1+x)]\Bigr) \Bigr\}.\label{so23}
\end{eqnarray}
We find analytically
\begin{equation}\label{so24}
\eta_{kk}=2k\ {\rm{Si}}(2k \pi).
\end{equation}
Eigen-solutions for  the   $2\times 2$ matrix have the form
\begin{eqnarray}
E_1=2.81019,\ E_3=5.99476 \ - \ {\rm{eigenvalues}}, \label{so25} \\
\psi (E_1)=(-0.995891,\ 0.0905574), \ \psi (E_3)=(0.0905574,\ 0.995891)
\ - \ {\rm{eigenvectors}}. \label{so26}.
\end{eqnarray}
Two lowest eigenvalues of   the  $6x6$ matrix  read $E_1=2.78021,\ E_3=5.93979$.
In Table  I  we reproduce the remaining four eigenvalues in the   $6\times 6$ case, in a comparative vein. Namely, we display
  the computation outcomes  for  lowest six  eigenvalues,  while  gradually increasing the matrix size,
  from  $6\times 6, 12\times 12, 5000\times 5000$ to  $10000\times 10000$.
    We reintroduce the traditional labeling in terms of $i=1,2,3,4,5,$   so that no
explicit distinction is made between even and odd eigenfunctions.
Our results are directly compared with the corresponding data obtained by other methods in Refs.  \cite{K,KKMS} and \cite{ZG,zg}.

In  Table II we report  the change of the  ground state energy while increasing the matrix size from $30\times 30$  to $10000\times 10000$.
 It is seen that the  third significant digit stabilizes already for $300\times 300$ and $400\times 400$ matrices.
\begin{table}
\begin{tabular}{|c|c|c|c|c|c|c|}
\hline
$i$ & 1 & 2 & 3 & 4 & 5 & 6 \\ \hline
Diagonal elem. &1.21531728 &2.83630315 & 4.38766562 & 5.96864490 & 7.53320446 & 9.10820377 \\ \hline
$E_{i6x6}$ &1.1704897 & 2.780209 & 4.356483317 & 5.9397942 & 7.52131594 & 9.099426 \\ \hline
$E_{i12x12}$ &1.1644016 &2.7690111 & 4.3388792 &5.919976& 7.4952827 &  9.0725254\\ \hline
$E_{i10^4x10^4}$ &1.157791 &2.754795 & 4.3168638 & 5.892233 & 7.460284 & 9.032984  \\ \hline
$E_{i(K)}$\cite{K} Table 2 &1.1577 & 2.7547 & 4.3168 & 5.8921 & 7.4601 & 9.0328 \\ \hline
$E_{i(KKMS)}$\cite{KKMS} Eq. (11.1)&1.1577738 & 2.7547547& 4.3168010& 5.8921474 & 7.4601757 &  9.0328526 \\ \hline
$E_{i(ZG)}$\cite{ZG} Table VII & 1.1560 & 2.7534  & 4.3168 & 5.8945& 7.4658 &  9.0427\\ \hline
$E_{i(zg)}$ \cite{zg} Table III & 1.157776 & 2.754769 & 4.316837 & 5.892214  & 7.460282 & * \\ \hline
\end{tabular}
\caption{Comparative table of 6 lowest eigenvalues $E_i$ in the  Cauchy  infinite potential well.  Results for
matrices of  different sizes in our approach  are compared with spectral data of  Refs. \cite{K}, \cite{KKMS}, \cite{ZG}
 and \cite{zg}. First six diagonal elements of the  matrix   (\ref{so13})  (expressions \eqref{so11} and \eqref{so24} respectively)
are cited for comparison.   Note that  the numbering of states  follows  tradition ($i=1,2,3,4,5,6$) and refers to
consecutive eigenvalues, with no reference   to the parity of respective  eigenfunctions. }
\end{table}

\begin{table}
\begin{tabular}{|c|c|c|c|c|c|c|c|c|c|}
\hline
$n$ (matrix $n\times n$)  & 30 & 50 & 100 & 200 & 400 &1000 &
2000 & 5000 & 10000 \\ \hline
$E_{g.s.}=E_1$ &1.160505 & 1.159428 & 1.158608 & 1.158193 & 1.157984 & 1.157858
& 1.157816 & 1.157791 &1.157791 \\ \hline
$E_2$ &2.760953 & 2.758572 & 2.756705 & 2.755742 & 2.755252 & 2.754954
& 2.754855 & 2.754795 &2.754795 \\ \hline
$E_3$ & 4.326418 & 4.322736 & 4.319842 & 4.318343 & 4.317578 & 4.317114
& 4.316958 & 4.316864 &4.316864 \\ \hline
$E_4$ &5.904768 & 5.900041 & 5.896238 & 5.894235 & 5.893204 & 5.892573
& 5.892361 & 5.892233 &5.892233\\ \hline
$E_5$ & 7.476052 & 7.470114 & 7.465334 & 7.462812 & 7.461511 & 7.460714
& 7.460446 & 7.460284 &7.460284 \\ \hline
$E_6$ &9.051406 & 9.044604 & 9.039015 & 9.036021 & 9.034462 & 9.033504
& 9.033180 & 9.032984 &9.032984 \\ \hline
\end{tabular}
\caption{The matrix $n \times n$-"size evolution"  of six lowest eigenvalues of \eqref{so13} as  $n$ grows.
 $E_{g.s.}$ stands for ground state energy. }
\end{table}

\subsection{Graphical comparison}

\begin{figure}[tbh]
\centering
\includegraphics[width=0.5\columnwidth]{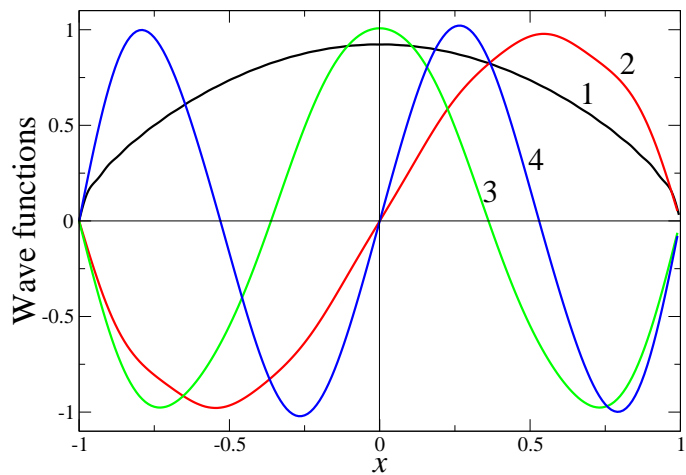}
\caption{Four lowest eigenfunctions in the   infinite Cauchy  well, labeled $i=1,2,3,4$. Outcome of  the  $10^4\times 10^4$ matrix.} \label{fig3}
\end{figure}

First, we plot the first four eigenfunctions  in Fig. \ref{fig3}. It is seen that qualitatively the states in the Cauchy  well  at a rough graphical resolution level do
resemble those (appear to be close) of  the ordinary quantum  infinite  well  (deriving form the Laplacian).  Anyway,  we know perfectly (see e.g. Section II) that  "plain"
trigonometric functions,  like e.g. $\cos(\pi x/2)$ or $\sin (\pi x)$,  are {\it  not} the  eigenfunctions  of  $|\Delta |^{1/2}_D$. \
Quite detailed analysis of the eigenfunctions shape issue can be found in Ref. \cite{zg}, where another method of solution of the Cauchy well  problem has been tested.

\begin{figure}[tbh]
\centering
\includegraphics[width=0.6\columnwidth]{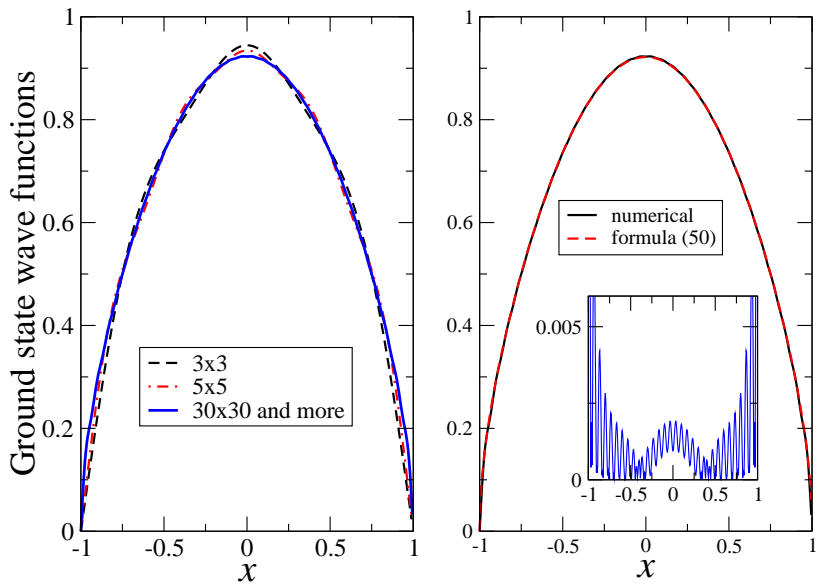}
\caption{Left panel. Comparison of the shapes of ground state  functions obtained by the diagonalization of 3x3 (black dashed curve), 5x5 (red dash-dot curve) and 30x30 (blue solid curve) matrices.
The shape of ground state functions for matrices more then 30x30 are identical to that for 30x30. Right panel shows the approximation of ground state wave function (for 700x700 matrix, solid curve)
 by the expression \eqref{zg} (dashed curve). As both lines are indistinguishable in the scale of the  figure, the inset depicts the modulus of the point-wise difference of respective curves} \label{fig4}
\end{figure}

Since, in the present paper,  we employ trigonometric functions as  the orthonormal basis  system,
  for low-sized matrices (\ref{so13})  we deal with  visually distinguishable oscillations.
These are gradually smoothened with the growth of the matrix size.
 It is instructive to  compare approximate  shapes of the  ground state function,
 obtained by the diagonalization of  different-sized  matrices.
The left panel of Fig. \ref{fig4} reports the  pertinent  shapes in case of  $3\times 3$, $5\times 5$ and $30\times 30$ matrices.
We note that the qualitative features of  the ground state function  approximants  are practically the same for matrices  of sizes exceeding
 $30\times 30$.

In Ref. \cite{zg} an analytic approximation of the ground state function  of $|\Delta |^{1/2}_D$  has been proposed   in the form
\begin{equation}\label{zg}
\psi_1(x)=\psi_{g.s.}(x)=0.921749\sqrt{(1-x^2)\cos \alpha x},\ \alpha =\frac{1443\pi}{4096}.
\end{equation}
In the right panel of Fig. \ref{fig4}   we  compare the ground state function \eqref{zg} with that obtained by the
 diagonalization of $700\times 700$ matrix  (which turns out to be close  to that obtained by  means of the  $30\times 30$ matrix,
 see Fig. \ref{fig5} below).
It is seen that both functions are indistinguishable within the scale of the figure. The inset in Fig. \ref{fig4}  depicts
 the modulus of the point-wise difference of these functions. Interestingly, although the approximation is non monotonous
 (the difference oscillates),  in a  large portion of f the interval $-1\leq x\leq 1$ the difference does not exceed $0.005$.

\subsection{Eigenvalues of $|\Delta |^{1/2}_D$.}

If compared with the previous methods of solution \cite{KKMS,K} and \cite{ZG,zg}, our spectral approach seems to be particularly powerful
if one is interested in the eigenvalues of $|\Delta |^{1/2}_D$. In fact, we are able to generate an  arbitrary number of
 eigenvalues, with a very high accuracy.
 In Table III we compare several (first 20 and  a couple of larger) lowest eigenvalues of $|\Delta |^{1/2}_D$
 and answer how much actually  the "rough" approximate formula  $n\pi /2 - \pi /8$  deviates from  computed  $E_n$s.
 That is motivated by the upper bound formula, \cite{K,KKMS} (in our notation and for the Cauchy stability  index $\alpha=1$),
 whose right-hand-side  drops  down to $0$ with
 $n\rightarrow \infty $:  $|E_n   -  \frac{n\pi}{2}+ \frac{\pi}{8} |  < \frac{1}{n}$.

\begin{table}
\begin{tabular}{|c|c|c|c|c|}
\hline
$n$ & $E_{n, 5000x5000}$ & $\frac{n\pi}{2}-\frac{\pi}{8}$ & Relative error, \% & Data from \cite{KKMS} \\ \hline
1 &1.157791 & 1.178097 & 1.75 & 1.157773 \\ \hline
2 & 2.754795 & 2.748894 & 0.21 & 2.754754 \\ \hline
3 & 4.316864 & 4.319690 & 0.06 & 4.316801   \\ \hline
4 & 5.892233 & 5.890486 & 0.03 &  5.892147 \\ \hline
5 & 7.460284 & 7.461283 & 0.013 &  7.460175  \\ \hline
6 & 9.032984 &  9.032079 & 0.01 &   9.032852 \\ \hline
7 & 10.602447 & 10.602875 & 0.004 & 10.602293   \\ \hline
8 & 12.174295 & 12.173672 & 0.0051 & 12.174118 \\ \hline
9 & 13.744308 & 13.744468 & 0.0012 &  13.744109 \\ \hline
10 & 15.315777 & 15.315264 & 0.0033 &  15.315554 \\ \hline
11 & 16.886062 & 16.886061 & 5.9$\cdot 10^{-8}$ & * \\ \hline
12 & 18.457329 & 18.456857 & 0.0026 & * \\ \hline
13 & 20.027767 & 20.027653 & 0.00057& * \\ \hline
14 & 21.598914 & 21.598449 & 0.0021 & *\\ \hline
15 & 23.169448 & 23.169246 & 0.00087 & * \\ \hline
16 & 24.740517 & 24.740042 & 0.0019  &  * \\ \hline
17 & 26.311115 & 26.310838 & 0.0011 &  * \\ \hline
18 & 27.882131 & 27.881635 & 0.0018 &  * \\ \hline
19 & 29.452773 & 29.452431 & 0.0012 &  * \\ \hline
20 & 31.023751 & 31.023227 & 0.0016 &  * \\ \hline
30 & 46.731898 & 46.731191 & 0.0015 &  * \\ \hline
50 & 78.148251 & 78.147117 & 0.0015 & * \\ \hline
100 & 156.689159 & 156.686934 & 0.0014 & * \\ \hline
\end{tabular}
\caption{Several lowest eigenvalues of the $5000 \times 5000$ matrix (\ref{so13}) are  presented.
  For comparison  the  approximate formula  $n\pi /2 - \pi /8$ is depicted together with the relative error
   $|E_n- (n\pi /2 - \pi /8)|/E_n$. Independently obtained spectral  data (formula 1.11 in \cite{KKMS})
   are displayed as well.}
\end{table}

It is seen from the Table III that although the asymptotic formula delivers pretty good approximation to the desirable
eigenvalues, the relative error never (except for $n=11$)  falls below $10^{-3}$ \%
as the label  number $n$ grows.  We  have actually   traced  this statement  up to $n= 500$.
Moreover, the relative error, as it is seen from the Table III, oscillates around $10^{-3}$ \%, which means that beginning
with $n \approx 8$ the expression $n\pi /2 - \pi /8$ contributes  5 significant digits of the "true" asymptotic  answer.\\

{\bf Technical comment:} We note here that to diagonalize large matrices ($30\times 30$  and larger)
we use the  Fortran program, based on the  LAPACK package. All  integrations involved in the evaluation
 of $\gamma_{ki}$ and $\eta_{ki}$
have been  performed numerically. \\

\section{Conclusions}

In the present paper we have elaborated a novel, independent from previous proposals, method of an  approximate
solution of the spectral problem of the infinite Cauchy well. Our method is based on the reduction of the
 initial spectral problem for the operator $-|\Delta|^{1/2}$ to the   Fredholm-type  integral equation with the hypersingular
 kernel. This equation, in turn, can be solved by  means of  the  (Fourier series)  expansion with respect  the
  complete set of orthogonal functions on the interval $-1\leq x \leq 1$ (trigonometric functions which are
   eigenfunctions of the Laplacian).

The adopted (Fourier series)  expansion method  transforms  the integral  eigenevalue problem  \eqref{ii2}
  to the eigenvalue  problem for an  infinite matrix.  We solve the approximate eigenvalue problems for  finite \
   matrices,  of the gradually increasing size. With the growth of the matrix size, new higher eigenvalues  are
   generated,  while  lower  eigenvalues becomes more and more accurate.
   We demonstrate that the lowest eigenfunctions  can  be approximately inferred by means of the  diagonalization of  relatively
   small  matrices, like e.g. $30\times 30$.
    We have noticed that the diagonal elements of  an approximating (finite) matrix give already good approximations for
    the  eigenvalues, see Table I.
To obtain the eigenvalues with $6$ significant digits the diagonalization of matrices  of the size $10000\times 10000$ of more
 is necessary.

The method appears to be a particularly powerful tool to compute the eigenvalues. It  can be generalized to
other  fractional (L\'{e}vy stable) operators, like e.g.    $-|\Delta |_D^{\mu /2}$,  $\mu \in (0,2)$.

\end{document}